\documentclass[sigconf]{acmart}

\usepackage{booktabs}
\usepackage{graphicx}
\newcommand{\covOneCityNearMe}{96}
\newcommand{\covOtherImplicit}{31}
\newcommand{\covOtherNearMe}{8}
\newcommand{\covTwinImplicit}{34}
\newcommand{\covTwinNearMe}{32}
\newcommand{\covUSCities}{15}
\newcommand{\curveFiveExplicit}{36}

\newcommand{\curveFiveNonlocal}{53}
\newcommand{\curveOneExplicit}{9}

\newcommand{\curveOneNonlocal}{19}

\newcommand{\curveTenExplicit}{69}

\newcommand{\defAnswersNamePlace}{11}
\newcommand{\defAsksNatLoc}{4}
\newcommand{\defAsksNatNone}{63}

\newcommand{\defJOther}{0.10}

\newcommand{\defJUS}{0.25}

\newcommand{\distMinKm}{293}
\newcommand{\distNearMeClose}{0.13}
\newcommand{\distNearMeFar}{0.14}
\newcommand{\distRho}{0.01}
\newcommand{\gemCountryInQueriesExplicit}{56}
\newcommand{\gemCountryInQueriesImplicit}{59}
\newcommand{\gemCountryInQueriesNonlocal}{4}
\newcommand{\gemDriftExplicit}{0.03}
\newcommand{\gemDriftImplicit}{0.01}
\newcommand{\gemDriftNonlocal}{0.05}
\newcommand{\gemNetCountryExplicit}{0.43}
\newcommand{\gemNetCountryExplicitCI}{0.40 to 0.46}
\newcommand{\gemNetCountryImplicit}{0.40}

\newcommand{\gemNetCountryNonlocal}{0.23}
\newcommand{\gemNetCountryNonlocalCI}{0.20 to 0.27}
\newcommand{\gemNetSameCountryExplicit}{0.00}
\newcommand{\gemNetSameCountryExplicitCI}{$-$0.03 to 0.04}

\newcommand{\gptBackendImplicit}{0.17}
\newcommand{\gptBackendImplicitCI}{0.09 to 0.25}
\newcommand{\gptDeclCityExplicit}{0.14}
\newcommand{\gptDeclCityImplicit}{0.17}
\newcommand{\gptDeclCityNonlocal}{0.00}
\newcommand{\gptDeclCountryExplicit}{0.28}
\newcommand{\gptDeclCountryImplicit}{0.43}
\newcommand{\gptDeclCountryNonlocal}{0.03}

\newcommand{\gptNetCountryExplicit}{0.00}

\newcommand{\gptNetCountryImplicit}{0.01}

\newcommand{\gptNetCountryNonlocal}{0.01}

\newcommand{\gptOwnPlaceOtherExplicit}{92}
\newcommand{\gptOwnPlaceOtherImplicit}{77}
\newcommand{\gptOwnPlaceOtherNonlocal}{10}
\newcommand{\gptOwnPlaceUSExplicit}{79}
\newcommand{\gptOwnPlaceUSImplicit}{22}
\newcommand{\gptOwnPlaceUSNonlocal}{1}
\newcommand{\gptPlaceInQueriesExplicit}{100}
\newcommand{\gptPlaceInQueriesImplicit}{48}
\newcommand{\gptPlaceInQueriesNonlocal}{0}
\newcommand{\gptRewriteImplicit}{0.37}
\newcommand{\gptSearchesExplicit}{100}
\newcommand{\gptSearchesImplicit}{67}
\newcommand{\gptSearchesNonlocal}{49}
\newcommand{\largeAnswers}{720}
\newcommand{\largeIdxExplicit}{0.26}
\newcommand{\largeIdxExplicitCI}{0.22 to 0.31}
\newcommand{\largeIdxImplicit}{0.35}

\newcommand{\largeIdxNonlocal}{0.11}

\newcommand{\largeLocations}{6}
\newcommand{\largePlaceInQueriesExplicit}{100}

\newcommand{\largeQuestions}{60}

\newcommand{\recBetweenNearMe}{4}
\newcommand{\recGemNetCountryExplicit}{0.41}

\newcommand{\recGemNetCountryImplicit}{0.28}

\newcommand{\recGemNetCountryNonlocal}{0.07}
\newcommand{\recGemNetCountryNonlocalCI}{0.04 to 0.10}
\newcommand{\recGptDeclCityExplicit}{0.18}

\newcommand{\recGptDeclCityImplicit}{0.14}

\newcommand{\recGptDeclCityNonlocal}{0.01}

\newcommand{\recGptDeclCountryExplicit}{0.22}

\newcommand{\recGptDeclCountryImplicit}{0.22}

\newcommand{\recGptDeclCountryNonlocal}{0.03}

\newcommand{\recGptNetCountryExplicit}{0.03}

\newcommand{\recGptNetCountryImplicit}{0.02}

\newcommand{\recGptNetCountryNonlocal}{0.00}

\newcommand{\recNamesPerNearMe}{4.3}
\newcommand{\recTwinNearMe}{23}
\newcommand{\searchArmIdxExplicit}{0.19}
\newcommand{\searchArmIdxImplicit}{0.20}
\newcommand{\searchArmIdxNonlocal}{0.01}
\newcommand{\smallIdxExplicit}{0.23}

\newcommand{\smallIdxImplicit}{0.31}

\newcommand{\smallIdxNonlocal}{0.07}

\newcommand{\topicLowImplicit}{0.02}
\newcommand{\topicLowName}{telecom, utilities}
\newcommand{\topicTopImplicit}{0.57}
\newcommand{\topicTopName}{public services}

\graphicspath{{figures/}}

\setcopyright{none}

\begin{document}

\title{Location, Location, Generation: Where You Ask Matters for Generative Engine Optimization}

\author{Mahe Chen}
\affiliation{\institution{University of Toronto}\country{Canada}}
\email{mahe.chen@mail.utoronto.ca}
\author{Nick Koudas}
\affiliation{\institution{university of Toronto}\country{Canada}}
\email{koudas@cs.toronto.edu}

\begin{abstract}
Generative search engines answer local and commercial questions with a written answer and a few cited
sources. A growing practice of generative engine optimization (GEO) monitors which sources and brands these
answers show, typically from a single place. In this paper, we present a controlled audit of how the user's
location changes the sources that generative engines cite. We ask 300 questions of ChatGPT with 19 declared
locations, and of ChatGPT and Gemini from five network locations, each twice to measure run-to-run noise
(31,200 answers). The engines take location from different channels. ChatGPT follows the location an
application declares, down to the city: it writes the place into its own searches, and its search backend
localizes results even when the queries name no place. It ignores the network. Gemini ignores the declared
location and follows the country of the network, changing most of its sources even for questions that do not
depend on place. The businesses the answers recommend change with the sources. Thus, visibility in generative
search is specific to the user's location, and audits must control both channels.
\end{abstract}



\maketitle

\section{Introduction}\label{sec:intro}

Generative search engines now answer many of the questions people used to type into a search box. Users ask
ChatGPT and Gemini where to find a plumber, which internet provider to choose or how much a car registration
costs. The engines answer with a text and a handful of cited sources, and these sources, together with
the businesses the answer names, decide what the user sees. An industry of generative engine optimization
(GEO)~\cite{aggarwal2024geo,chen2025geo} monitors these answers for brands and publishers, typically by sending a
fixed set of questions from one place.

For classic web search, location is the main source of personalization apart from being logged
in~\cite{hannak2013personalization}, and its effect varies with the kind of query~\cite{kliman2015location}.
Whether and how generative engines personalize by location has not been measured. This matters for three
reasons. A monitoring result obtained in one place may not hold elsewhere; a local business competes for
visibility within its location; and an audit must know which signal of location an engine uses before it can
control it.

We address three research questions. \textbf{RQ1:} Does the user's location change which sources a generative
engine cites, and for which kinds of questions? \textbf{RQ2:} Through which channel does location enter: a
location the application \emph{declares} to the engine, or one the engine \emph{infers} from the network; and,
for a declared location, the model writing the place into its searches or the search backend?
\textbf{RQ3:} What changes: local or national sources, and the businesses the answers recommend?

In this paper, we present a controlled location audit of generative search with a twin-run noise baseline. We
show that ChatGPT and Gemini take location from different channels, at different granularity, separate query
rewriting from backend localization, and derive implications for GEO monitoring and audits.

\section{Related Work}\label{sec:related}

Audits of web search personalization compare result pages for the same query under controlled conditions and
measure the difference against the noise between identical requests~\cite{hannak2013personalization}. Location
changes results more for local queries~\cite{kliman2015location}. Audits of generative search study whether
cited sources support the answer~\cite{liu2023verifiability}, how engines select and absorb sources across
platforms~\cite{zhang2026absorption}, and how Gemini and AI Overviews relate to Google
Search~\cite{grossman2026disrupts}. Chen et al.~\cite{chen2025geo} compare AI search with Google and find that it
favors earned media over brand-owned and social content, and that the engines differ in domain
diversity, freshness, cross-language stability and sensitivity to phrasing; we add location as a further
dimension. Generative engines retrieve sources with queries they write
themselves~\cite{lewis2020rag,ma2023rewriting}, which gives location a second route into the results. Work on
GEO evaluates content changes on fixed sets of sources~\cite{aggarwal2024geo,puerto2025cseo}. We are not aware
of any work that varies the user's location.

\section{Method}\label{sec:method}

\paragraph{Questions and locations.} We wrote 300 questions in ten commercial topics (home services, health
care, finance, insurance, telecom, dining, retail, cars, travel, public services), with 100 per group:
\emph{near me} questions that ask for something nearby (``best plumber near me''), questions whose answer
\emph{depends on the place} (prices, laws, providers, weather; ``how much is sales tax on a car''), and
\emph{not local} questions (``best robot vacuum''). No question names a place. The declared locations are 15
US cities that differ in size and region, and four cities in other English-speaking countries (London, Toronto,
Sydney, Dublin).

\paragraph{E1: declared location.} We send every question to ChatGPT (gpt-5.4-mini, Responses API with the
web search tool), twice, with each city as the tool's approximate user location and without a location. The
model decides by itself whether to search; a second arm prefixes the question with ``Search the web:'', so
that sources exist for every question. E1 comprises 24,000 answers, all sent from one network. A robustness run
repeats the first arm with the full-size model (gpt-5.4) on a random sample of 60 questions (20 per group) at
six locations.

\paragraph{E2: network location.} We send every question without a declared location to ChatGPT (``Search
the web:'' arm) and to Gemini (gemini-3.1-flash-lite with Google Search grounding), twice each, in six
sessions from different network locations: the authors' own connection in Canada, VPN exits in New York, Los
Angeles, London and Sydney, and the own connection again three days later to measure drift (7,200 answers).
Note that Gemini's API also accepts declared coordinates; in a pilot they had no effect, which E2 explains.

\paragraph{Measures.} For each answer, we record whether it searched, its search queries, its cited URLs, and
whether a query names its location. For a question, we compare the cited domains of two answers with the
Jaccard index. Twins (same question, setting and session) give the noise; two settings with the same repeat
give the difference. The \emph{localization index} is the mean twin overlap minus the mean overlap between
settings, computed per question and averaged over questions with a bootstrap 95\% confidence interval; 0
means that location has no effect beyond noise. A source is \emph{place-specific} if its URL names the
answer's city or US state, or carries the country's domain ending. To capture recommendations, a language
model lists the businesses, brands and products each answer recommends (directories, review sites and public
bodies excluded); in a hand check, 29 of 36 extractions were fully correct.

\section{Results}\label{sec:results}

\begin{figure}[t]
\centering
\includegraphics[width=\columnwidth]{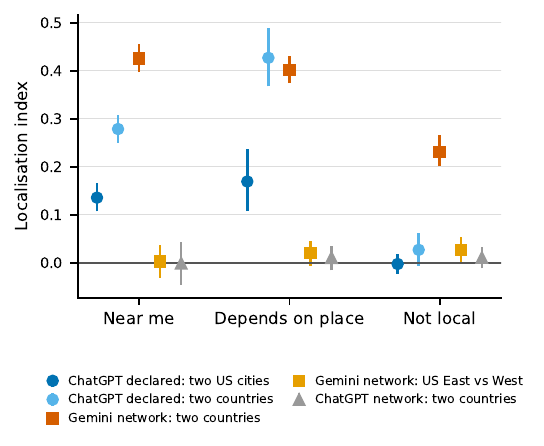}
\caption{Localization index of cited domains (twin overlap minus overlap between settings; 95\% CI over
questions). Declared: ChatGPT with a stated city (E1). Network: no stated location, different network
locations (E2).}
\Description{Dot plot of the localization index for three question groups. ChatGPT with declared locations
and Gemini with network locations in different countries show large positive values for near-me and
place-dependent questions; Gemini also for not-local questions. ChatGPT with network locations and Gemini
between US East and US West stay at zero.}
\label{fig:index}
\end{figure}

\begin{figure}[t]
\centering
\includegraphics[width=\columnwidth]{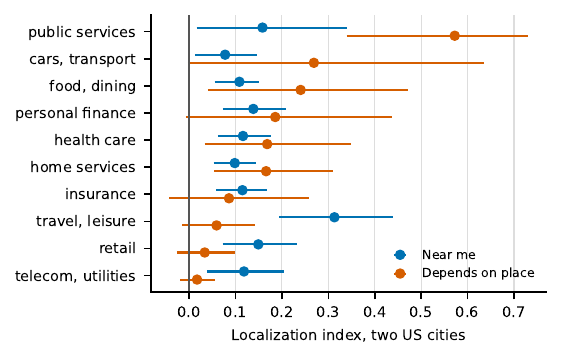}
\caption{Localization index of cited domains between two US cities by topic (ChatGPT, declared location;
95\% CI over the ten questions of each topic and group).}
\Description{Horizontal dot plot of the localization index by topic for near-me and place-dependent questions.
Public services have the highest index for place-dependent questions; retail and telecom are close to zero.}
\label{fig:topics}
\end{figure}

\paragraph{RQ1: Location changes the sources of local answers.} With a declared location
(Figure~\ref{fig:index}), ChatGPT's sources for near-me questions differ between two US cities by
\gptDeclCityExplicit{} and between cities in different countries by \gptDeclCountryExplicit; for
place-dependent questions, by \gptDeclCityImplicit{} and \gptDeclCountryImplicit. For not-local questions it
is \gptDeclCityNonlocal{} and \gptDeclCountryNonlocal{}; their sources do not depend on the declared location.
Gemini's sources differ between network locations in different countries by \gemNetCountryExplicit{}
(CI \gemNetCountryExplicitCI) for near-me questions and \gemNetCountryImplicit{} for place-dependent ones.
Interestingly, they still differ by \gemNetCountryNonlocal{} (CI \gemNetCountryNonlocalCI) for not-local ones. A
declared location also changes whether ChatGPT answers with sources at all. Without one, it often asks for a
ZIP code instead; with one, it cites sources in 64--73\% of its answers, against 46\% without. Localization
also differs by topic (Figure~\ref{fig:topics}). Among place-dependent questions, it is strongest for
\topicTopName{} (\topicTopImplicit), whose answers cite the government sites of each state, and weakest for
\topicLowName{} (\topicLowImplicit), whose answers cite national sources such as the FCC, national carriers and
comparison sites. With ten questions per topic and group, the topic estimates are indicative.

\paragraph{RQ2: The two engines use different channels.} ChatGPT ignores the network. Between network locations
in different countries, its index is \gptNetCountryExplicit, \gptNetCountryImplicit{} and
\gptNetCountryNonlocal, all within noise. Gemini ignores declared coordinates and follows the network, but only
at the level of the country. Between New York and Los Angeles its index is \gemNetSameCountryExplicit{}
(CI \gemNetSameCountryExplicitCI), and its queries name the country (\gemCountryInQueriesExplicit\% of near-me
answers) but never a city. Drift does not explain the country effect: the same connection three days apart
gives \gemDriftExplicit, \gemDriftImplicit{} and \gemDriftNonlocal{}, and London and Sydney, measured 30 minutes
apart, differ as much as any other pair of countries. Table~\ref{tab:queries} shows how each engine writes the
location into its searches. Within ChatGPT, location enters in two ways. First, the model writes the place into
\gptPlaceInQueriesExplicit\% of its searches for near-me questions and \gptPlaceInQueriesImplicit\% for
place-dependent ones.

\begin{table}[t]
\caption{A search query each engine wrote for the same question, by place (first of two requests).}
\label{tab:queries}
\small
\begin{tabular}{@{}lp{5.9cm}@{}}
\toprule
\multicolumn{2}{@{}l}{\textit{ChatGPT, declared: ``credit unions in my area''}} \\
\quad Boise & ``credit unions Boise Idaho area branches near Boise Idaho'' \\
\quad New York & ``credit unions near New York NY'' \\
\quad London & ``credit unions near London England United Kingdom'' \\
\addlinespace
\multicolumn{2}{@{}l}{\textit{Gemini, network: ``which store has the cheapest groceries''}} \\
\quad Canada & ``is it cheaper to shop at Walmart vs No Frills vs Superstore Canada'' \\
\quad US & ``cheapest grocery stores in US 2026'' \\
\quad UK & ``is Aldi or Lidl cheaper UK 2026'' \\
\quad Australia & ``how to save money on groceries Australia'' \\
\bottomrule
\end{tabular}

\end{table} Second, the search backend localizes by itself. Where neither of two answers' queries names its
location, the sources still differ (index \gptBackendImplicit, CI \gptBackendImplicitCI); for example, ChatGPT
cites Oregon State University's extension service on when to plant tomatoes for a user in Boise. Where both
queries name the place, the index rises to \gptRewriteImplicit{}. Moreover, ChatGPT's localization does not fade
with distance. For near-me questions, the index is \distNearMeClose{} between US cities less than 1,000 km apart
and \distNearMeFar{} between cities more than 3,000 km apart (Spearman correlation with distance \distRho; our
closest pair is \distMinKm{} km apart). Thus, to ChatGPT every other city is a different place; to Gemini, only
another country is.

\paragraph{The default location.} Without a declared location, ChatGPT answers \defAsksNatNone\% of near-me
questions by asking for a ZIP code or city (\defAsksNatLoc\% with a location). When it does search, it writes no
state into its queries, its sources resemble those of the US cities (overlap \defJUS) much more than those of
the other countries (\defJOther), and only \defAnswersNamePlace\% of its near-me answers name any US place. Thus,
it answers for an unspecified user in the United States, from national sources. Gemini's default is the
country of the network, whatever location is declared.

\begin{table}[t]
\caption{Behavior and place-specific sources by question group. ChatGPT with a declared location (E1, question
as typed); Gemini from five network locations (E2). Own-place sources: answers citing a URL that names the
answer's city or US state (US) or carries its country's domain ending (other).}
\label{tab:behaviour}
\small
\setlength{\tabcolsep}{3pt}
\begin{tabular}{@{}lrrrrr@{}}
\toprule
 & \multicolumn{4}{c}{ChatGPT, declared} & Gemini \\
\cmidrule(lr){2-5}\cmidrule(l){6-6}
 & Searches & Place in & \multicolumn{2}{c}{Own-place sources} & Country in \\
 & & queries & US & other & queries \\
\midrule
Near me & \gptSearchesExplicit\% & \gptPlaceInQueriesExplicit\% & \gptOwnPlaceUSExplicit\% & \gptOwnPlaceOtherExplicit\% & \gemCountryInQueriesExplicit\% \\
Place-dependent & \gptSearchesImplicit\% & \gptPlaceInQueriesImplicit\% & \gptOwnPlaceUSImplicit\% & \gptOwnPlaceOtherImplicit\% & \gemCountryInQueriesImplicit\% \\
Not local & \gptSearchesNonlocal\% & \gptPlaceInQueriesNonlocal\% & \gptOwnPlaceUSNonlocal\% & \gptOwnPlaceOtherNonlocal\% & \gemCountryInQueriesNonlocal\% \\
\bottomrule
\end{tabular}
\end{table}

\paragraph{RQ3: Local sources and different businesses.} Table~\ref{tab:behaviour} shows what changes. With a
declared location, \gptOwnPlaceUSExplicit\% of ChatGPT's near-me answers in US cities cite a source that names
the city or state, and \gptOwnPlaceOtherExplicit\% of those in other countries cite a source from that country;
a placebo that counts sources naming another of our locations stays below 0.2\%.

The businesses, brands and products the answers recommend change with the sources (Table~\ref{tab:recs}). A
near-me answer from ChatGPT names \recNamesPerNearMe{} of them on average. Two identical requests share
\recTwinNearMe\% of these names; two different declared locations share \recBetweenNearMe\%. For Gemini, the
recommended names differ between network countries even for not-local questions (index
\recGemNetCountryNonlocal, CI \recGemNetCountryNonlocalCI), whereas ChatGPT's recommendations do not change
with the network. Table~\ref{tab:examples} shows two typical questions. Asked for credit unions, ChatGPT names
local institutions in each declared city. Asked which store has the cheapest groceries, a question that names
no place, Gemini recommends different national chains in each network country, and the same chains in New
York and Los Angeles.

\begin{figure}[t]
\centering
\includegraphics[width=\columnwidth]{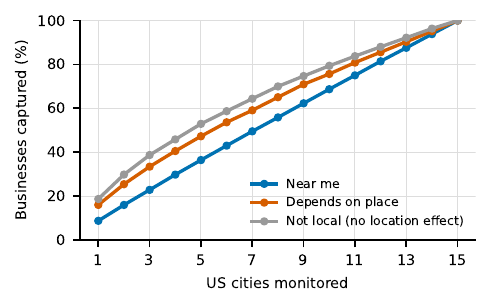}
\caption{Share of the businesses that ChatGPT recommends across 15 US cities (two requests each) captured by
monitoring $k$ randomly chosen cities (declared location, question as typed).}
\Description{Line chart with three rising lines from one to fifteen monitored cities. Near-me questions rise
almost linearly from 9 to 100 percent; place-dependent and not-local questions start higher and curve slightly.}
\label{fig:coverage}
\end{figure}

\paragraph{Implications for monitoring.} GEO tools track which businesses an engine recommends by asking
questions from one place. For near-me questions, a request declared in one US city recommends only
\covOtherNearMe\% of the businesses that the same request recommends in another of our \covUSCities{} cities,
against \covTwinNearMe\% when the request is repeated in the same city. Moreover, \covOneCityNearMe\% of the
businesses recommended for near-me questions appear in only one of the \covUSCities{} cities. For place-dependent
questions the gap is smaller (\covOtherImplicit\% against \covTwinImplicit\%). For near-me questions, one city
(Figure~\ref{fig:coverage}) captures \curveOneExplicit\% of the businesses
recommended across the \covUSCities{} cities, five cities \curveFiveExplicit\% and ten \curveTenExplicit\%. The
curve does not level off, since every added city brings mostly new businesses. Not-local questions, which
location does not affect, give the reference: one city captures \curveOneNonlocal\% and five
\curveFiveNonlocal\%, due to the variation between repeated requests. Thus, monitoring needs both repeated
requests and, for local questions, many locations.

\paragraph{Robustness.} In the second arm, where ChatGPT searches for every question, the localization index
is the same (\searchArmIdxExplicit,
\searchArmIdxImplicit{} and \searchArmIdxNonlocal{} over all location pairs). The full-size model, run on
\largeQuestions{} questions at \largeLocations{} locations (\largeAnswers{} answers), localizes at least as
strongly as the small one. Its index is \largeIdxExplicit{} (CI \largeIdxExplicitCI), \largeIdxImplicit{} and
\largeIdxNonlocal{}, against \smallIdxExplicit, \smallIdxImplicit{} and \smallIdxNonlocal{} for the small model
on the same questions and locations, and it writes the place into \largePlaceInQueriesExplicit\% of its near-me
searches.

\begin{table}[t]
\caption{Localization index of the recommended businesses, brands and products (twin overlap minus overlap
between settings). Values with 95\% CIs are in the repository.}
\label{tab:recs}
\small
\setlength{\tabcolsep}{4pt}
\begin{tabular}{@{}lrrr@{}}
\toprule
 & Near me & Place-dep. & Not local \\
\midrule
ChatGPT, declared: two US cities & \recGptDeclCityExplicit & \recGptDeclCityImplicit & \recGptDeclCityNonlocal \\
ChatGPT, declared: two countries & \recGptDeclCountryExplicit & \recGptDeclCountryImplicit & \recGptDeclCountryNonlocal \\
Gemini, network: two countries & \recGemNetCountryExplicit & \recGemNetCountryImplicit & \recGemNetCountryNonlocal \\
ChatGPT, network: two countries & \recGptNetCountryExplicit & \recGptNetCountryImplicit & \recGptNetCountryNonlocal \\
\bottomrule
\end{tabular}
\end{table}

\begin{table}[t]
\caption{The first three businesses recommended in both twin answers to the same question, by place.}
\label{tab:examples}
\small
\begin{tabular}{@{}lp{5.6cm}@{}}
\toprule
\multicolumn{2}{@{}l}{\textit{ChatGPT, declared: ``credit unions in my area''}} \\
\quad New York & New York University FCU, Neighborhood Trust FCU, Brooklyn Cooperative FCU \\
\quad Boise & Idaho Central CU, CapEd FCU, Connections CU \\
\quad Miami & Educational FCU, University CU, Miami Postal Service CU \\
\quad London & London Mutual CU, London Capital CU, Your Community Bank \\
\addlinespace
\multicolumn{2}{@{}l}{\textit{Gemini, network: ``which store has the cheapest groceries''}} \\
\quad Canada & No Frills, Walmart, Food Basics \\
\quad US East & Aldi, Walmart, Costco \\
\quad US West & Aldi, Walmart, WinCo Foods \\
\quad UK & Aldi, Lidl, Tesco \\
\quad Australia & ALDI, Coles, Woolworths \\
\bottomrule
\end{tabular}

\end{table}

\section{Discussion and Limitations}\label{sec:discussion}

\paragraph{What the two channels mean for GEO.} Visibility in generative search is not one number: it depends
on where the user is, and on which signal of location the engine reads. Three consequences follow for
monitoring. First, the location must be set on the channel the engine uses: a declared location for ChatGPT, a
network location for Gemini. A tool that sets only a location parameter measures ChatGPT for that city but
Gemini for the country it runs from; a tool that sets neither measures ChatGPT for an unspecified US user and
Gemini for its own country. Second, the granularity differs. Since ChatGPT's localization does not fade with
distance, no city stands in for another, not even a nearby one; every market a business serves must be
measured on its own. For Gemini, one network location per country suffices, but a brand present in several
countries must be measured in each, even for questions that do not mention a place. Third, comparisons over
time or between engines are valid only with the location held fixed, since changing the location changes most
of the sources and recommended businesses.

For optimization, the results suggest where visibility is decided. For near-me and place-dependent questions,
ChatGPT writes the place into its searches and cites sources that name the city or state. As such, a local
business becomes visible through pages that the engine finds for place-specific queries, such as local
directories, review sites and the sites of local institutions, rather than through national rankings. For
Gemini, the country of the user decides which national chains and sources are recommended; a brand's visibility
in one country says little about another. The same holds for audits of generative engines: any study run
through Gemini's API is localized by the network it runs from, unless that location is controlled.

\paragraph{Limitations.} We use the engines' APIs and small models, which need not behave like the consumer
applications; the full-size model check covers ChatGPT only. VPN exits are data-center addresses and cover five
network locations; Gemini may localize more finely for residential connections. The questions are in English
and were written by us. The measurements describe the engines in October 2026.

\section{Conclusion}\label{sec:conclusion}

Motivated by single-place monitoring of generative search, we examined how the user's location changes the
sources that generative engines cite.
ChatGPT followed the location an application declares, down to the city, through its own queries and its search
backend; Gemini followed the country of the network, even for questions that do not depend on place. Thus,
monitoring and auditing generative engines must fix both the declared and the network location. We release the
questions, code and answers.


\bibliographystyle{ACM-Reference-Format}
\bibliography{references}

\end{document}